\documentclass[amsmath,amssymb,aps,prb,floatfix, twocolumn, 10pt, longbibliography]{revtex4-2}
\usepackage[x11names, svgnames, dvipsnames]{xcolor}
\usepackage{bbm}
\definecolor{maroon}{RGB}{128,0,0}
\usepackage{hyperref}
\hypersetup{colorlinks,linkcolor={maroon},citecolor={maroon},urlcolor={maroon}}
\usepackage{graphicx}
\usepackage{dcolumn}
\usepackage{bm}
\usepackage{comment}
\usepackage{braket}
\usepackage{textcomp, gensymb}
\usepackage{silence}
\begin{document}

\title{Disorder-induced superconductivity in graphene}

\author{Jannes van Poppelen}
\email{jannes.vanpoppelen@physics.uu.se}
\affiliation{Department of Physics and Astronomy, Uppsala University, Box 524, SE-751 20 Uppsala, Sweden
}

\author{Tomas Löthman}
\affiliation{Department of Physics and Astronomy, Uppsala University, Box 524, SE-751 20 Uppsala, Sweden
}

\author{Annica M. Black-Schaffer}
\affiliation{Department of Physics and Astronomy, Uppsala University, Box 524, SE-751 20 Uppsala, Sweden
}

\date{\today}

\begin{abstract}
Correlated phases of matter are typically investigated in crystalline systems, where disorder is considered to be detrimental. However, intriguing exceptions exist, such as superconductivity being enhanced in amorphous realizations of Al and Bi. Here, we demonstrate that superconductivity can even emerge entirely from disorder, using monolayer graphene as an example. In the clean limit, the semi-metallic nature of graphene requires prohibitively strong electronic interactions to achieve superconductivity. Despite the inherently random nature of disorder, we show that introducing low concentrations of vacancies or hydrogenation in graphene provides a large density of low-energy states that easily induce superconductivity.
For conventional $s$-wave pairing, disorder induces a finite superconducting order parameter for arbitrarily weak attractive interactions. Rather than forming isolated superconducting puddles, global phase coherence is established through a finite superfluid weight of purely geometrical origin. Away from the chiral limit of vacancies, hydrogenation similarly yields a finite transition temperature and nonzero superfluid weight for weak interactions. For unconventional nearest-neighbor pairing, typically more disrupted by disorder, superconductivity exhibits quantum-critical-like behavior, yet phase coherence persists at low interaction strengths, with mixed $d$-wave symmetries. Our work demonstrates the robust emergence of macroscopic superconducting phase coherence engineered entirely from microscopic disorder.
\end{abstract}

\maketitle
Despite graphene’s exceptional tunability, intrinsic superconductivity in pristine monolayers remains experimentally undetected due to a lack of low-energy states, although theoretical predictions exist for high doping \cite{nandkishore2012chiral, black2014chiral}. In contrast, the emergence of moir\'e physics has established twisted multilayer graphene as a highly versatile class of superconductors \cite{cao2018unconventional, yankowitz2019tuning, torma2022superconductivity}, driven by the formation of zero-energy twist-induced flat bands \cite{bistritzer2011moire, lisi2021observation}. More recently, the realization of intrinsic superconductivity has expanded beyond twisted graphene to rhombohedral multilayer graphene, where the ABC-stacking generates flat bands at its surfaces, tunable with electric displacement fields \cite{zhou2021superconductivity, han2025signatures}. These findings firmly establish the presence of effective attractive interactions for superconductivity in graphene systems.

Random vacancies offer an alternative method for generating zero-energy states in graphene. To first approximation, vacancies preserve chiral symmetry, and is therefore one class of chiral disorder, that avoid wave-function localization at any disorder concentration and generate a singular density of states (DOS) at low energy \cite{gade1993anderson, pereira2006disorder, ostrovsky2006electron, stauber2007electronic, wehling2010resonant, ugeda2010missing, ostrovsky2014density}. However, the disorder renders the electronic structure highly inhomogeneous and fractious. A fundamental question thus arises as to whether such highly disordered samples can sustain a globally phase coherent superconducting state?

Recent developments in quantum geometry have clarified how flat-band systems, such as moir\'e materials, achieve superconducting phase coherence, despite the absence of electron dispersion. This coherence arises from a geometric contribution to the superfluid weight, which depends on the normal-state quantum metric \cite{kopnin2011surface, peotta2015superfluidity, julku2020superfluid, peotta2025quantum, yu2025quantum}, the real part of the quantum geometric tensor. Intriguingly, it has recently been shown that vacancy pairs in graphene generate a finite quantum metric, with rare, long-distance pairs contributing significantly \cite{marsal2024enhanced}.

Here, we demonstrate how chiral disorder in graphene generates a globally phase-coherent superconducting state, even at vanishingly small attractive interactions. Relaxing the chiral symmetry by using experimentally easily accessible hydrogenation \cite{gonzalez2016atomic} similarly gives rise to superconductivity at weak interactions. First, by assuming conventional $s$-wave superconductivity, we show that even small concentrations of disorder enable a finite superconducting order parameter at arbitrarily weak interaction strengths. Then, we establish that global phase coherence emerges as a direct consequence of disorder-induced quantum geometry, which we quantify by computing the superfluid weight and the Berezinskii–Kosterlitz–Thouless (BKT) transition temperature, $T_{\text{BKT}}$. Phase coherence is maximized around 10 \% disorder concentration, due to a diminishing quantum metric for larger concentrations.
We further also consider unconventional nearest-neighbor pairing, thereby covering the two lowest harmonic spin-singlet pairing channels. By drawing analogies to the Lieb lattice, we demonstrate quantum critical point (QCP)-like behavior, while maintaining phase coherence for weak interactions.
The disorder generates a real-valued mixture of primarily the $d_{xy}$- and $d_{x^2-y^2}$-wave superconducting symmetries. Although disorder is generally disruptive to unconventional pairing, we still find it induces superconductivity. 
Our results establish superconductivity entirely driven by disorder and also provide a mechanism for finally achieving superconductivity in monolayer graphene.

\begin{figure*}
    \centering
    \includegraphics[width=\linewidth]{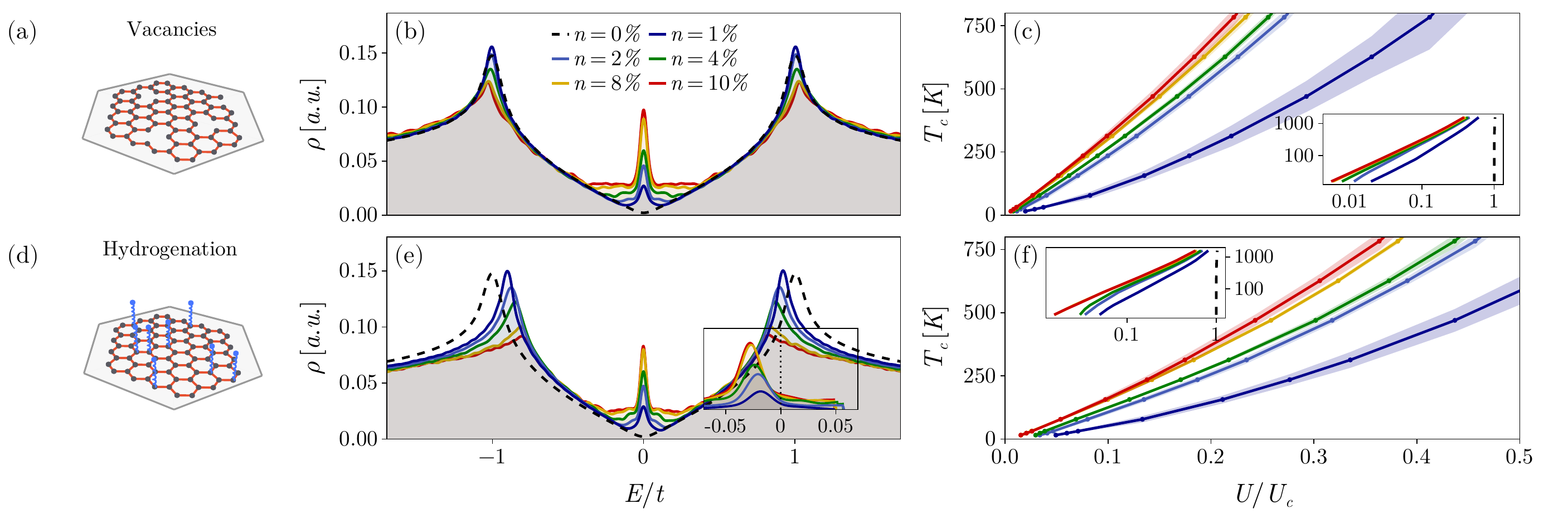}
    \caption{\textbf{Normal-state spectrum and superconducting $T_c$ of disordered graphene.} (a,d) Schematic lattice realizations with  (a) vacancies and (d) hydrogenation, with carbon atoms (grey dots) with nearest neighbor bonds (red), and adatoms with bonds (blue). (b,e) Normal-state density of states $\rho$ as a function of energy $E$ for varying disorder concentration $n$, using a Gaussian energy smearing of $0.02\,t$. Inset in (e) highlights the chiral symmetry breaking induced by hydrogenation, resolved with finer smearing $0.01\,t$. (c,f) Superconducting pairing critical temperature $T_c$ as a function of on-site interaction strength $U$ normalized to the critical $U_c$ for superconducting pairing in pristine graphene. Insets include $T_c$ for pristine graphene (dashed black line). Results are averaged over five disorder configurations. Shaded regions indicate standard deviation of the mean.}
    \label{fig:fig1}
\end{figure*}

\section*{Disordered graphene}
We describe graphene using a tight-binding Hamiltonian characterized by a nearest-neighbor hopping $t$ and first consider a random distribution of vacancies (Methods and Fig.~\ref{fig:fig1}(a)), which induce a pronounced low-energy peak in the DOS, $\rho$, that scales with the vacancy concentration $n$ (Fig.~\ref{fig:fig1}(b)). In contrast to the van Hove singularity peaks at high energy, the chiral-disorder DOS peak is constrained to zero energy by chiral symmetry \cite{gade1993anderson, ostrovsky2014density}. Consequently, electrostatic gating is not required to align the Fermi level with the DOS singularity and it can instead serve as a tool to finely tune the system's properties around the chiral peak.

To assess robustness beyond the chiral limit upheld by vacancies, we also model randomly distributed hydrogen adatoms \cite{wehling2010resonant}, which hybridize with the carbon lattice and carry a finite on-site energy (Methods and Fig.~\ref{fig:fig1}(d)). This breaks the chiral symmetry and shifts the resonant energy away from charge neutrality. Nevertheless, a high degree of chiral symmetry is retained \cite{usaj2014partial}. Hydrogenation results in a comparable enhancement of the low-energy DOS, albeit centered at a finite resonance energy (inset in Fig.~\ref{fig:fig1}(e)).

\section*{Disorder-induced pairing instability}
Having established that disorder generates a flat-band-like DOS, we next explore the possibility of superconductivity arising from these low-energy defect states. Notably, with superconductivity now found to be prevalent in a multitude of flat-band graphene systems \cite{cao2018unconventional, yankowitz2019tuning, zhou2021superconductivity, han2025signatures}, the necessary effective attractive pairing interaction for superconductivity is also most likely present in monolayer graphene.
We start by modeling conventional spin-singlet $s$-wave pairing via an on-site Hubbard attraction of strength $U$, modeling phonon-driven superconductivity (Methods).

To identify the onset of pairing, we employ the linear gap equation (LGE) to determine the transition temperature $T_c$ for both vacancies and hydrogenation (Methods and Figs.~\ref{fig:fig1}(c,f)), setting the Fermi level at the disorder-induced DOS peak. Unlike conventional superconductors, where a finite DOS enables pairing at arbitrarily weak attraction, pristine graphene possesses a vanishing DOS at charge neutrality and thereby requires a large critical interaction $U_c \approx 2.23\,t$ \cite{black2009effect} for superconductivity (black dashed line in inset in Figs.~\ref{fig:fig1}(c)). We find that both types of disorder qualitatively alter this behavior, even at low concentrations $n$, by drastically reducing the required interaction strength. Indeed, vanishingly small interactions produce a finite $T_c$, although hydrogenation leads to a somewhat lower $T_c$ compared to vacancies due to the shifted and lower DOS peak.
Crucially, both disorder types yield a linear scaling $T_c \sim |U|$, a hallmark of flat-band superconductivity \cite{kopnin2011high, peotta2025quantum}, which contrasts sharply with the exponential BCS result.

\begin{figure}
    \centering
    \includegraphics[width=1\columnwidth]{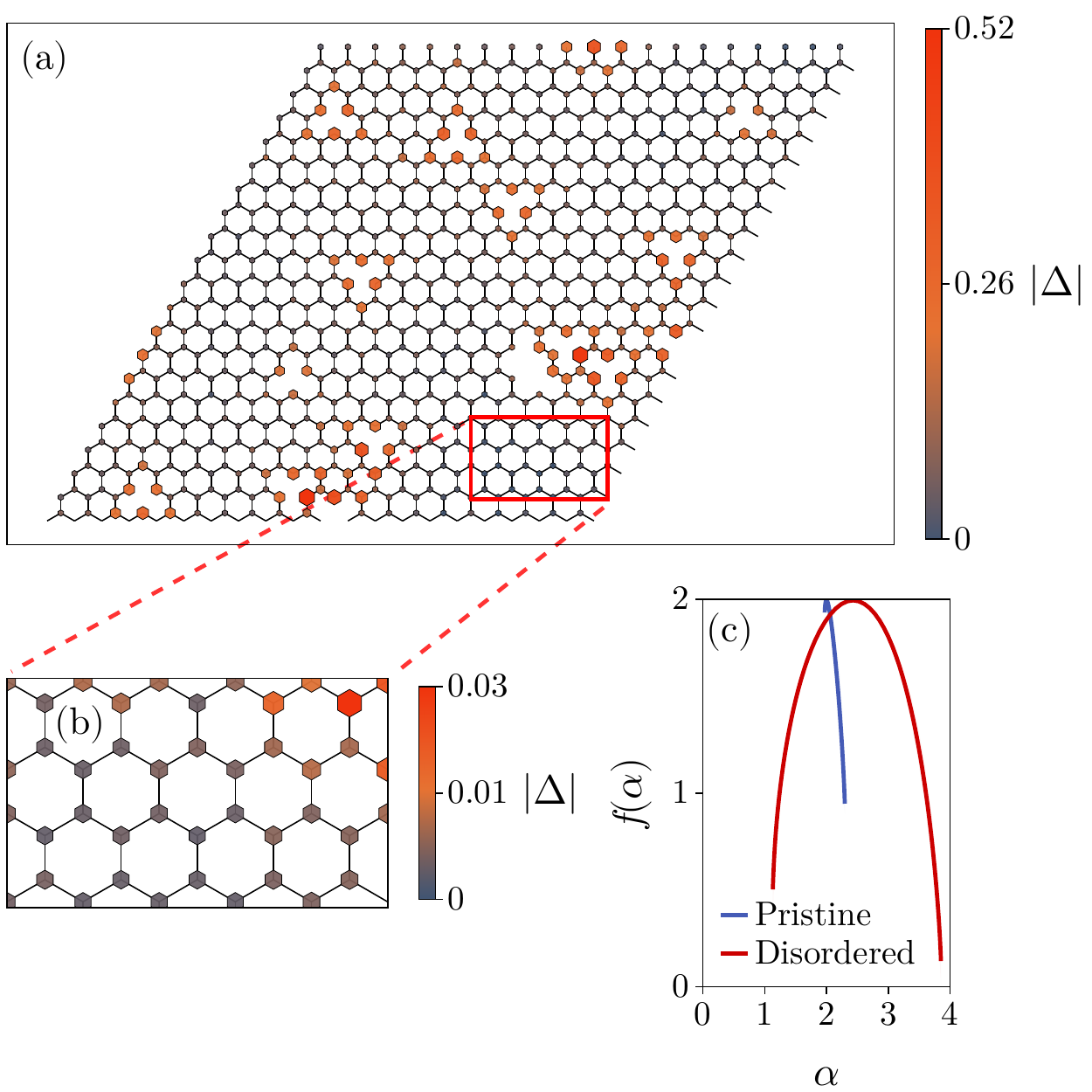}
    \caption{\textbf{Spatial inhomogeneity and multifractality of the superconducting state.} (a) Real space map of magnitude of on-site superconducting order parameter $|\Delta|$ for disorder concentration $n=4\,\%$ at interaction strength $U=0.67\,U_c$. (b) Zoom-in of superconducting order parameter in a pristine region. (c) Multifractal singularity spectrum $f(\alpha)$ of $|\Delta|$ for $n=8\,\%$, averaged over five disorder configurations. }
    \label{fig:fig2}
\end{figure}

To characterize the superconducting order, we self-consistently solve the non-linear gap equation (NLGE, Methods) at $T=0$, for a low vacancy concentration and interaction well below the quantum critical point (QCP) of pristine graphene. The spatially resolved order parameter $\Delta$ (Fig.~\ref{fig:fig2}(a)) reveals a highly inhomogeneous state, with pairing primarily localized around vacancy sites. This localization is driven by the underlying low-energy defect state, which reside on the sublattice opposite to the vacancy. Although seemingly vanishing in pristine regions, pairing is present throughout the sample, albeit one order of magnitude smaller (Fig.~\ref{fig:fig2}(b)), which we attribute to the critically localized, multifractal defect states, which decay algebraically rather than exponentially \cite{evers2008anderson}.
Hydrogenation yields a similar spatial profile (Extended Data Fig.~\ref{fig:s1}) but with a lower magnitude, in agreement with its lower low-energy DOS peak (Figs.~\ref{fig:fig1}(b,e)).

We find that the superconducting order parameter directly reflects the multifractality of the defect states. The critical nature of these defect states implies that a continuum of scaling exponents $\alpha$ is necessary to describe the algebraic scaling of their wave functions, characterized by a singularity spectrum $f(\alpha)$ \cite{chhabra1989direct, lopes2009fractal}. We therefore compute and compare $f(\alpha)$ between a uniform and a self-consistent order parameter on the disordered vacancy lattice (Methods and Fig.~\ref{fig:fig2}(c)). While the uniform order parameter yields only the trivial spatial dimension ($f(\alpha=2)=2$), the self-consistent solution spans a broad range of $\alpha$. This explicitly demonstrates that the spatial profile of the order parameter inherits the critical and multifractal character of the defect states, beyond the inhomogeneity of the disordered lattice.

\section*{Phase coherence and superfluid weight} 
Given the highly inhomogeneous nature of the superconducting pairing, it is crucial to also determine whether global, long-range phase coherence can be sustained, necessary for a superconducting condensate.
The emergence of a global superconducting state in two dimensions is ultimately prevented by quantum phase fluctuations. Phase coherent superconductivity can still manifest through topological order rather than conventional symmetry breaking, via the BKT mechanism \cite{berezinskii1971destruction, kosterlitz1973ordering}. The associated transition temperature $T_{\text{BKT}}$ is dictated by the superfluid weight tensor $D^{s}$, which characterizes the energy cost of phase fluctuations. At $T_{\text{BKT}}$, $D^s$ undergoes a universal jump that marks the onset of vortex unbinding, defining the Nelson-Kosterlitz criterion \cite{kosterlitz1974critical, nelson1977universal}.

In conventional superconductors, the superfluid weight is driven entirely by the band dispersion in the normal state. However, recent advances in quantum geometry have established that the superfluid weight also contains a geometric contribution arising from the quantum metric of the normal state \cite{peotta2015superfluidity, peotta2025quantum, yu2025quantum}. The total superfluid weight naturally decomposes into these two distinct physical components, $D^s = D^s_{\text{conv}} + D^s_{\text{geom}}$ \cite{liang2017band, huhtinen2022revisiting, lau2022universal, lamponen2025superconductivity}. In flat-band systems the conventional term $D^s_{\text{conv}}$ vanishes, but the geometric part $D^s_{\text{geom}}$ can still provide a finite superfluid weight.

\begin{figure}
    \centering
    \includegraphics[width=\columnwidth]{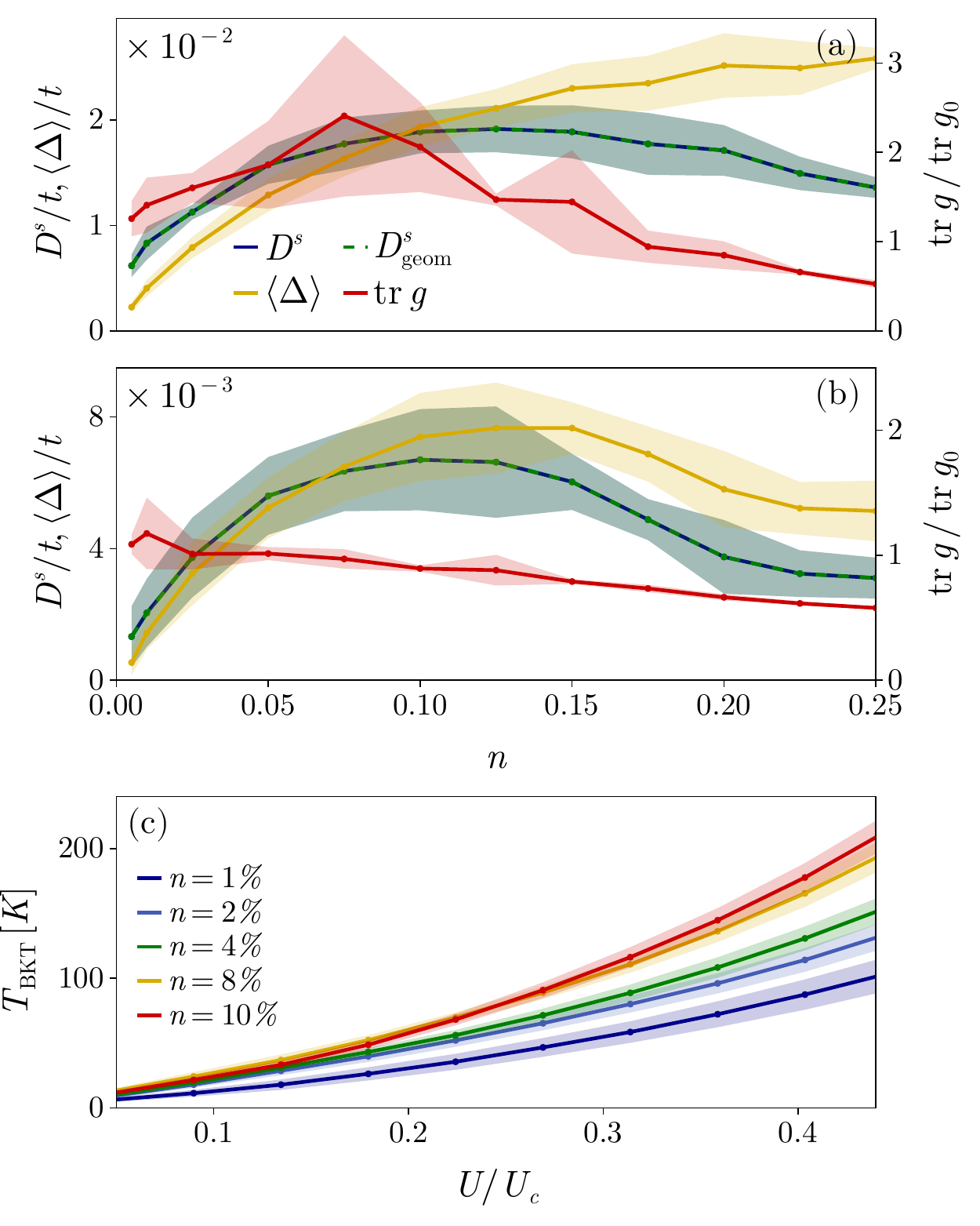}
    \caption{\textbf{Disorder-induced superfluid weight and BKT temperature.} (a,b) Zero-temperature superfluid weight $D^s$, its geometric contribution $D^s_{\text{geom}}$, average superconducting order parameter $\langle\Delta\rangle$ (left axis), and trace of the normal-state quantum metric $\mathrm{tr}\,g$ (right axis, normalized to the pristine graphene quantum metric $\mathrm{tr}\,g_0$), as a function of disorder concentration $n$ at interaction strength $U=0.45\, U_c$ for vacancy disorder (a) and hydrogenation (b), averaged over five disorder realizations. Note the smaller scale in (b) compared to (a). (c) BKT temperature $T_{\text{BKT}}$ as a function of interaction strength $U$ for different vacancy concentrations, averaged over three disorder realizations. Shaded regions indicate standard deviation of the mean.}
    \label{fig:fig3}
\end{figure}

To quantify the ability for a phase-coherent superconducting state, we compute the spatially averaged superconducting order parameter $\langle\Delta\rangle$, the trace of the normal-state quantum metric $g$, and the zero-temperature superfluid weight $D^s$ within linear response (Methods) for random vacancies and hydrogenation (Figs.~\ref{fig:fig3}(a,b)), well within the superconducting pairing regime. We find that even a small concentration of vacancies or hydrogenation enables a finite superfluid weight. Crucially, we find that the conventional contribution to $D^s$ is zero, which we attribute to the low-energy DOS peak being entirely disorder induced, akin to a disorder-induced flat band. Consequently, the emergence of a phase-coherent state becomes entirely dependent on the quantum geometry. Interestingly, in pristine and undoped graphene, the superconducting weight is of equal conventional and geometrical origin \cite{kopnin2008bcs, kopnin2010supercurrent, liang2017band}. The purely geometric component in disordered graphene must therefore arise as a direct consequence of the disorder. Thus, we find that disorder does not only generate the zero-energy DOS peak necessary for pairing but also enhances the quantum geometry required for phase coherence.

For vacancy disorder (Fig.~\ref{fig:fig3}(a)), the trace of the quantum metric $g$ is significantly enhanced with disorder, peaking at $n\approx7\,\%$, in excellent agreement with earlier work \cite{marsal2024enhanced}. Interestingly, the superfluid weight $D^s$ reaches its maximum at a slightly higher vacancy concentration, $n\approx12\,\%$. This occurs because the superfluid weight is not determined solely by the quantum metric. Specifically, for an isolated flat band, the relation $D^s \propto |\Delta|\int \text{tr}\,g(\boldsymbol{k})\,\mathrm{d}\boldsymbol{k}$, with the integral over the Brillouin zone \cite{peotta2015superfluidity, liang2017band, huhtinen2022revisiting}, shows that $D^s$ depends also on the superconducting order parameter $\Delta$. Although the disorder-induced flat band in disordered graphene is not perfectly isolated, $D^s$ interpolates between the quantum metric and the order parameter, peaking at a higher concentration than the quantum metric. 

For hydrogenation (Fig.~\ref{fig:fig3}(b)), we find a lower total superfluid weight $D^s$ compared to vacancy disorder. 
We can attribute this to two effects. First, the induced quantum metric from hydrogenation is significantly lower. In fact, the quantum metric has been shown to be very sensitive to the chiral symmetry breaking \cite{marsal2024enhanced} present for hydrogenation.
Second, while the presence of chiral symmetry for vacancies ensures that the system remains at half-filling, hydrogenation introduces local potential fluctuations that drive the system away from charge neutrality. This increased charge inhomogeneity suppresses the superconducting state at larger disorder concentrations, causing $\langle\Delta\rangle$ to peak at around $n=15\%$, instead of monotonically increasing as observed for vacancy disorder. 

To further assess the superconducting state, we compute $T_{\mathrm{BKT}}$ using the Nelson-Kosterlitz criterion (Methods) for vacancy disorder. The results (Fig.~\ref{fig:fig3}(c)) demonstrate that phase coherence emerges at temperatures significantly lower than the superconducting order parameter (Fig.~\ref{fig:fig1}(c)). Despite this difference, the superconducting transition remains remarkably robust, reaching values of tens of kelvins even for very small $U$.
Notably, $T_{\mathrm{BKT}}$ also increases with disorder concentration, reflecting the disorder-induced quantum geometry.
In the weak-coupling limit ($U\ll t$), the transition temperature exhibits linear scaling $T_{\text{BKT}} \sim|U|$. For larger $U$, however, $T_{\text{BKT}}$ begins to deviate from this linear behavior, akin to the behavior induced by higher-energy bands in flat band systems \cite{peotta2025quantum}.

For hydrogenation, we instead compute the zero-temperature superfluid weight as function of disorder concentration (Extended Data Fig.~\ref{fig:s3}), due to the increased computational cost of self-consistently determining $\mu$ to align the disorder-induced DOS peak with the Fermi energy. Similarly to vacancy disorder, we find a finite superfluid weight over a broad range of interaction strengths, showcasing that phase coherence is similarly supported. We here attribute the apparent onset at finite $U$, as well as the non-monotonic behavior in $n$ at large $U$, to the spatially inhomogeneous Hartree shift due to chiral symmetry breaking.
Taken together, our results demonstrate that disorder-enhanced zero-energy DOS and quantum geometry establishes a robust and global phase coherent conventional superconducting state in graphene, even at minimal attractive interactions.

\section*{Unconventional pairing}
\begin{figure*}
    \centering
    \includegraphics[width=\linewidth]{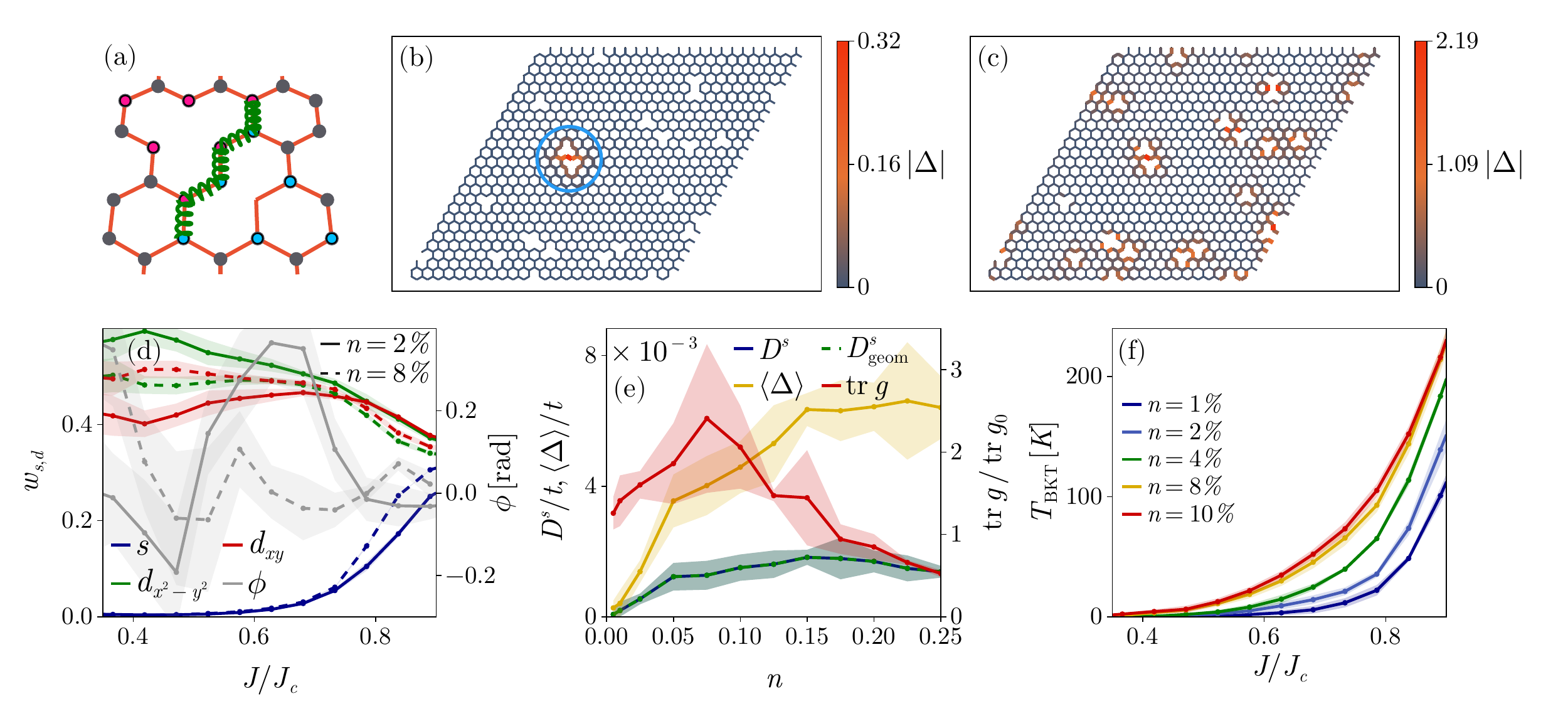}
    \caption{\textbf{Unconventional superconductivity.} (a) Illustration of nearest-neighbor pairing from vacancies. Pink and blue markers represent sites with the most weight of the two vacancy-induced defect states, and green helices represent dominant bond pairing. (b,c) Spatially resolved unconventional superconducting order parameter for vacancy concentration $n=2\,\%$ at interaction strength $J =0.52\,J_c$ (b) and $J =0.84\,J_c$ (c). Blue circle highlights the bond pairing illustrated in (a). (d) Pairing symmetry resolved weight ($w_s$, $w_{d_{x^2-y^2}}$, $w_{d_{xy}}$, left axis) and relative phase between $d$-wave components ($\phi$, right axis) of the superconducting order parameter as a function of $J$ for two different vacancy concentrations. (e) Superfluid weight $D^s$, geometric contribution $D^s_{\text{geom}}$, average superconducting order parameter $\langle\Delta\rangle$ (left axis), and trace of the normal-state quantum metric $\mathrm{tr}\,g$ (right axis, normalized to the pristine graphene quantum metric $\mathrm{tr}\,g_0$) as a function of vacancy concentration at interaction strength $J=0.58\, J_c$. Results in (d,e) are averaged over five disorder configurations. (f) BKT temperature $T_{\text{BKT}}$ for as a function of interaction strength $J$ for different vacancy concentrations, averaged over three disorder configurations. Shaded regions in (d-f) indicate standard deviation of the mean.}
    \label{fig:fig4}
\end{figure*}

Although conventional pairing provides foundational understanding of superconductivity in disordered graphene, it neglects the role of non-local electronic correlations. We therefore extend our analysis to include unconventional pairing through modeling spin-singlet pairing on nearest-neighbor bonds with interaction strength $J$ (Methods). Such pairing appears both within the strong-coupling $t$-$J$ limit of the repulsive Hubbard model and through spin fluctuations, as extensively studied in heavily doped graphene  \cite{black2007resonating, Kiesel2012competing, black2012chiral, nandkishore2012chiral}. 

We first examine the possibility for unconventional superconducting pairing. As with the conventional $s$-wave pairing, pristine graphene hosts superconductivity only beyond a QCP at $J_c\approx1.91\,t$ \cite{black2007resonating}. For disordered graphene, we illustrate a minimal example of nearest-neighbor pairing using a pair of vacancies placed on different sublattices (Fig.~\ref{fig:fig4}(a)), with pink and blue markers indicating the sites with most weight for the two zero-energy defect states and green helices representing the dominant pairing. The long-range algebraic decay of the critically localized vacancy states facilitates pairing from opposite-sublattice vacancies also at large separations. However, a single  vacancy fails to induce finite nearest-neighbor pairing since its defect state resides exclusively on one sublattice. Due to these restrictions we find that a QCP-like behavior still exists for disorder-driven unconventional pairing, but at an interaction strength much lower than for pristine graphene. We note that this behavior mirrors that of the Lieb lattice, where zero-energy compact localized states exhibit the same single-sublattice support and superconductivity likewise only emerges beyond a QCP \cite{lamponen2025superconductivity}. 
Since any unconventional pairing will have a non-local spatial structure, we conclude already here that unconventional superconductivity is more restricted than conventional superconductivity in disordered graphene.

We next solve the NLGE to characterize the resulting superconducting order at zero temperature (Methods) and plot the real-space amplitude of the superconducting order parameter. At moderate interactions ($J =0.52\,J_c$), the amplitude is strongly localized at double-vacancy sites (blue circle, Fig.~\ref{fig:fig4}(b))), consistent with our prediction (Fig.~\ref{fig:fig4}(a)). 
At this disorder concentration, we observe a strongly suppressed order parameter for lower interaction strengths, reflecting the QCP-like behavior. However, increasing disorder concentration lowers the interaction threshold at which the order parameter vanishes.
As we further increase the coupling to $J = 0.84\,J_c$ (Fig.~\ref{fig:fig4}(c)), the superconducting state becomes more robust. Now, it also emerges around single vacancies, as well as around divacancies on the same sublattice, highlighting the effect of a random distribution of critically localized defect states.

To further understand the superconducting state, we decompose the local order parameter into the irreducible representations of the honeycomb lattice (Methods, Fig.~\ref{fig:fig4}(d)). We find that the spatially averaged order parameter is dominated by a mixture of the two $d$-wave channels (red, green), supplemented by a small extended $s$-wave component (blue). In heavily doped graphene, the chiral $d_{x^2-y^2} + i d_{xy}$ state is favored \cite{black2014chiral}. Here, we find that disorder breaks the degeneracy of the $d$-wave channels and also suppresses the relative phase between them from the chiral $\pi/2$ value (grey). This displays a transition towards disorder-driven nematicity, though the relative phase shows no clear monotonic dependence on disorder concentration or interaction strength. As we approach $J_c$ all three pairing symmetries begin to equally contribute, as expected for pristine undoped graphene \cite{black2007resonating}.

We further compute the spatially averaged superconducting order parameter $\langle\Delta\rangle$, the trace of the normal-state quantum metric $g$, and the superfluid weight $D^s$ for nearest-neighbor pairing (Fig.~\ref{fig:fig4}(e)). We observe no significant differences in these quantities compared to conventional pairing (Fig.~\ref{fig:fig3}(a)), although the superfluid weight for nearest-neighbor pairing peaks at a slightly higher disorder concentration.
Finally, we compute the BKT transition temperature $T_\mathrm{BKT}$ (Fig.~\ref{fig:fig4}(f)) and find that the behavior differs noticeably from that of conventional pairing (Fig.~\ref{fig:fig3}(c)).
For all disorder concentrations, we observe an apparent onset of phase-coherent superconductivity only beyond a finite interaction threshold, which decreases as the vacancy concentration increases.
Combined with and observed exponential scaling, this strongly suggest QCP-like behavior.
Furthermore, the resulting $T_\mathrm{BKT}$ values are smaller than those obtained for conventional pairing when compared through their relative interaction strengths, i.e.~$J/J_c$ and $U/U_c$.  Nevertheless, over a broad range of interaction strengths, disorder induces phase-coherent superconductivity well before pristine graphene becomes superconducting.
 
We also compare the superconducting order parameter for hydrogenation to that of vacancy disorder (Extended Data Fig.~\ref{fig:s2}). Although the spatial profiles are qualitatively similar, the magnitude is smaller for hydrogenation. This is expected, as both types of defect states are strongly sublattice-polarized, while the lower DOS peak in the hydrogenated system suppresses the superconducting state.

\section*{Discussion}  
We have demonstrated that disorder in graphene can give rise to a highly inhomogeneous, yet phase-coherent superconducting state, even for vanishingly small interactions. As a multitude of graphene-systems are now known to be superconducting, this offers great possibilities for developing entirely disorder-driven superconductivity. 
While our investigation have focused on spin-singlet pairing, our results can easily be generalized to spin-triplet pairing. Nearest-neighbor $p$-wave pairing should behave similarly to our results for nearest neighbor spin-singlet pairing. We therefore expect QCP-like behavior but at a substantially lower critical interaction strength than pristine graphene. The main difference is that no extended $s$-wave channel can emerge, possibly promoting a highly nematic superconducting state. 
Going beyond nearest neighbor pairing, spin-triplet $f$-wave pairing also becomes possible, along with higher-order $p$- and $d$-wave harmonics \cite{pangburn2023superconductivity}. Here, the key is the sublattice pairing structure, since vacancy (hydrogenation) defect states have a perfect (strong) sublattice polarization. Inter-sublattice pairing should behave similar to the nearest-neighbor pairing results in Fig.~\ref{fig:fig4}. Intra-sublattice pairing, such as on next nearest neighbor bonds, should combine the intra-sublattice features of on-site pairing in Fig.~\ref{fig:fig2} with the directional dependence of nearest neighbor pairing in Fig.~\ref{fig:fig4}. In particular, this opens up a pathway toward disorder-induced and fully gapped spin-triplet $f$-wave pairing that is not locked behind a QCP, as in clean or in few layer ABC-stacked graphene \cite{awoga2023superconductivity}.

The results in this work are derived within the idealized limit where vacancies in graphene possess exact chiral symmetry. The inclusion of longer-range hopping and intra-orbital hybridization weakly breaks both the chiral and particle-hole symmetries \cite{castro2009electronic}. Indeed, first-principles calculations indicate that the bandwidth of vacancy-induced defect states is not strictly zero, but instead broadens to up to 0.5~eV \cite{de2022vacancy}. While this is around 3-4 times larger than the width of the DOS peaks in Fig.~\ref{fig:fig1}, this increased energy spread will only suppress the superconducting transition temperature if the interaction strength is smaller than the bandwidth of the defect states.

Finally, enhanced zero-energy DOS may also drive other emergent electronic orders in disordered graphene. It has been demonstrated that both vacancies and hydrogenation in graphene generate local magnetic moments \cite{palacios2008vacancy, gonzalez2016atomic}. Fortunately, vacancy-induced magnetic moments are inherently unstable due to structural lattice reconstruction \cite{lee2005diffusion, ozccelik2013self} and those arising from hydrogenation are highly sensitive to electrostatic gating and readily vanish with small doping \cite{nair2013dual}. We therefore propose slight doping to induce superconducting orders free from competition with magnetic orders.
Indeed, superconductivity has been shown to generically appear as domes flanking  magnetic order \cite{lothman2016universal} when doping away from a flat band, a phenomenon also clearly observed in systems ranging from twisted \cite{cao2018unconventional, yankowitz2019tuning} to rhombohedral multilayer graphene \cite{zhou2021superconductivity, han2025signatures}.

\nocite{*}

\clearpage
\section*{Methods}
\subsection*{Normal state}
We consider a graphene lattice $\mathcal{G}$ with $N$ sites, with nearest-neighbor hopping $t$. Vacancies are introduced by removing a set of sites $\mathcal{V}$, corresponding to a vacancy concentration $n = |\mathcal{V}|/N$. Any sites that become dangling as a result are subsequently removed iteratively. At higher vacancy concentrations, this pruning procedure removes an increasing number of dangling sites and can cause the effective vacancy concentration to notably exceed $n$. For example, a starting concentration of $n = 10\,\%$ yields an effective, reported, vacancy concentration of approximately $15\,\%$.

The resulting Hamiltonian is
\begin{equation}
    \label{eq:hamiltonian} 
    H_0 = -t\!\!\!\!\!\!\!\sum_{\sigma,\,\langle i,j\rangle \in \mathcal{G}\backslash\mathcal{V}}\!\!\!\!\!\!\!\!c_i^{\dagger}c_j - \mu\!\!\!\!\sum_{\sigma,\,i\in \mathcal{G}\backslash\mathcal{V}}\!\!\!\!\!c^{\dagger}_ic_i,
\end{equation}
where $t=2.7$ eV and $\mu$ is the chemical potential \cite{castro2009electronic}. 

For hydrogenation, the randomly distributed adatoms each add an extra orbital with energy $\epsilon$ which hybridize with the pristine graphene lattice ($\mathcal{V}=\emptyset$) with a strength $V$. In this case, the total non-interacting Hamiltonian becomes $H_0 \rightarrow H_0 + H_{\text{ad}}$, where
\begin{equation}
\label{eq:adham}
    H_{\text{ad}} = \epsilon\sum_j d_j^{\dagger}d_j + V\,c_j^{\dagger}d_j + \mathrm{H.c.}
\end{equation}
Matching $H_0$ to ab-initio calculations gives $\epsilon = -t/16$ and $V = 2t$ \cite{wehling2010resonant}. 

We perform calculations on graphene supercells containing 5000 carbon sites before introducing vacancies, which we verify to be free of finite-size effects. Periodic boundary conditions are imposed to eliminate spurious zero-energy edge states. For visualization, we plot fewer than 5000 sites.

\subsection*{Superconducting pairing}
\subsubsection*{Conventional pairing}
To model conventional spin-singlet $s$-wave pairing we use an attractive Hubbard interaction ($U>0$), acting  on each carbon atom resulting in
\begin{equation}
    \label{eq:schamiltonian}
    H = H_0 - U\sum_{i\in \mathcal{G}\backslash\mathcal{V}} n_{i\uparrow}n_{i\downarrow}.
\end{equation}
Within a mean-field treatment, the Hamiltonian becomes
\begin{equation}
    \label{eq:hamiltonianMF} 
    H_{\text{MF}} = H_0- \sum_{\sigma,\,i\in \mathcal{G}\backslash\mathcal{V}}\mu_ic^{\dagger}_ic_i + \sum_{i\in \mathcal{G}\backslash\mathcal{V}}\Delta_ic^{\dagger}_{i\uparrow}c^{\dagger}_{i\downarrow} +{\rm H.c.},
\end{equation}
with the mean-field order parameters $\Delta_i = U\langle c_{i\uparrow}c_{i\downarrow}\rangle$ and $\langle n_i\rangle=\sum_{\sigma}\langle c^{\dagger}_{i\sigma}c_{i\sigma}\rangle$. The local chemical potential $\mu_i$ is thereby Hartree shifted as $\mu_i = \mu+\frac{U}{2}\langle n_i\rangle$. In pristine graphene, translational symmetry enforces a uniform $\mu_i$. Disorder, however, generates spatially localized low-energy states, leading to strong inhomogeneity in $\mu_i$.

To determine the superconducting state, we solve self-consistently for the order parameter $\Delta_i = U\langle c_{i\uparrow}c_{i\downarrow}\rangle$ together with the local densities $\langle n_i\rangle = \sum_{\sigma}\langle c^{\dagger}_{i\sigma}c_{i\sigma}\rangle$, within the Bogoliubov-de Gennes formalism (BdG) \cite{zhu2016bogoliubov}. For this, we transform to the supercell momentum space and we introduce the Nambu spinor $\Psi_{\boldsymbol{k}} = \begin{pmatrix} \{c_{\boldsymbol{k}}\}, & \{c_{-\boldsymbol{k}}^{\dagger}\}\end{pmatrix}^T$, where the site indices are omitted for brevity. The mean-field Hamiltonian is then given by $H_{\text{MF}} = \sum_{\boldsymbol{k}} \Psi_{\boldsymbol{k}}^{\dagger} H_{\text{BdG}}(\boldsymbol{k}) \Psi_{\boldsymbol{k}}$, with the BdG Hamiltonian
\begin{equation}
\label{eq:bdg}
    H_{\text{BdG}}(\boldsymbol{k}) =\begin{pmatrix}
        H_0(\boldsymbol{k}) & \Delta(\boldsymbol{k})\\
        \Delta^{\dagger}(\boldsymbol{k}) & -H_0^*(-\boldsymbol{k})
    \end{pmatrix}.
\end{equation}

For each momentum $\boldsymbol{k}$, $H_{\text{BdG}}$ is diagonalized via a Bogoliubov transformation, yielding eigenenergies $\{E_{n\boldsymbol{k}}\}$ and eigenstates $|\psi^n_{\boldsymbol{k}}\rangle$ for the $n$-th state. Projecting these states onto the local basis yields the quasiparticle amplitudes $\{\psi^n_{\boldsymbol{k}}=\begin{pmatrix} u^n_{\boldsymbol{k}} & v^n_{\boldsymbol{k}} \end{pmatrix}^T\}$. The resulting superconducting order parameters and local densities at each carbon site are then computed as
\begin{equation}
\begin{aligned}
\label{eq:nlge}
\Delta_i &= \frac{U}{2N_k} \sum_{n\boldsymbol{k}} u_{i\boldsymbol{k}}^{n}v_{i\boldsymbol{k}}^{n*}\tanh\left(\frac{E_{n\boldsymbol{k}}}{2T}\right), \\
\langle n_i \rangle &= \frac{1}{N_k} \sum_{n\boldsymbol{k}} \left[ |u_{i\boldsymbol{k}}^{n}|^2 f(E_{n\boldsymbol{k}}) + |v_{i\boldsymbol{k}}^{n}|^2 (1 - f(E_{n\boldsymbol{k}})) \right],
\end{aligned}
\end{equation}
which constitute the non-linear gap (NLGE) and density equations, respectively. Here, $f$ is the Fermi–Dirac distribution, $T$ is the temperature, and $N_k$ is the number of $\boldsymbol{k}$-points in the supercell Brillouin zone (BZ), which we sample using the method of Monkhorst and Pack \cite{monkhorst1976special}. We select $N_k$ such that we achieve good convergence of the DOS, which typically results in $N_k =$ 4 to 9.

We determine the superconducting order parameter $\Delta_i$ at $T=0$ by solving the NLGE self-consistently. Starting from an initial guess, we iteratively diagonalize the BdG Hamiltonian in Eq.~\eqref{eq:bdg} and update the order parameter $\Delta_i$ via Eq.~\eqref{eq:nlge} until the relative difference between consecutive iterations falls below $10^{-4}$. 
For hydrogenation, we further also solve the density equation in Eq.~\eqref{eq:nlge} self-consistently for $\mu$, such that the disorder-induced DOS peak aligns with the Fermi energy within a tolerance of $5\cdot 10^{-2}\,\text{eV}$. For vacancy disorder, we can instead simply fix $\mu = -U/2$ to maintain half-filling, as it locks the disorder-induced flat band to the Fermi energy. This simplification is a consequence of the uniform density theorem for bipartite systems \cite{lieb1993uniform}, which guarantees half-filling under particle-hole symmetry.
To handle the high computational cost of the exact diagonalization of the BdG Hamiltonian, we leverage GPU-accelerated routines as implemented in Julia \cite{bezanson2017julia, besard2018juliagpu}.

\subsubsection*{Unconventional pairing}
We also consider unconventional pairing in the form of spin-singlet nearest-neighbor bond superconductivity arising from spin fluctuations or within the $t$-$J$ model \cite{black2007resonating, Kiesel2012competing, black2012chiral, nandkishore2012chiral}. The mean-field Hamiltonian for this interaction becomes
\begin{equation}
    \label{eq:hamiltonianMFNN} 
    H_{\text{MF}} = H_0 + \sum_{\langle i,j\rangle\in \mathcal{G}\backslash\mathcal{V}}\Delta_{ij}\left(c^{\dagger}_{i\uparrow}c^{\dagger}_{j\downarrow} - c^{\dagger}_{i\downarrow}c^{\dagger}_{j\uparrow}\right)+ {\rm H.c.},
\end{equation}
where the superconducting order parameter $\Delta_{ij} = J\langle c_{i\downarrow}c_{j\uparrow}-c_{i\uparrow}c_{j\downarrow}\rangle$ now resides on the bonds linking atoms $i$ and $j$ and with an interaction strength $J$. The only other distinction from the on-site pairing case is the absence of a Hartree shift in the chemical potential \cite{black2007resonating}. Within the BdG formalism, these superconducting order parameters are computed as
\begin{equation}
\Delta_{ij} = \frac{J}{2N_k}\sum_{n\boldsymbol{k}}u_{i\boldsymbol{k}}^{n}v_{j\boldsymbol{k}}^{n*}\tanh\left(\frac{E_{n\boldsymbol{k}}}{2T}\right).
\end{equation} 

\subsubsection*{Pairing symmetry decomposition}
We can decompose the local pairing amplitudes $\Delta_{ij}$ into the irreducible representations (irreps) of the honeycomb lattice with point group symmetry $D_{6h}$. For conventional pairing, only $s$-wave symmetry is allowed.
For nearest neighbor spin-singlet pairing, the point group decomposes into a one-dimensional extended $s$-wave channel ($A_{1g}$ representation) and a two-dimensional, degenerate $d$-wave symmetry channel ($E_{2g}$ representation) \cite{black2007resonating, black2014chiral}. Their orthonormal basis vectors, on the three nearest neighbor bonds are $\phi_s = \frac{1}{\sqrt{3}}(1, 1, 1)$, $\phi_{d_{x^2-y^2}} = \frac{1}{\sqrt{6}}(2, -1,-1)$, and $\phi_{d_{xy}} = \frac{1}{\sqrt{2}}(0, 1,-1)$.
We express the bond order parameter as $\Delta_{ij} = \sum_{\chi} \Delta^{\chi}_{ij}$, where the components $\Delta^{\chi}_{ij}$ are obtained by projecting $\Delta_{ij}$ onto these basis vectors $\phi_\chi$ for $\chi \in \{s, d_{x^2-y^2}, d_{xy}\}$.
We restrict ourselves to sites that have three intact bonds. Missing bonds due to vacancies explicitly break the local point group symmetry, destroy the orthogonality of the basis vectors and artificially mix distinct pairing channels \cite{canyellas2026extended}.

Because the superconducting state is highly inhomogeneous, we report a weighted global average to characterize the dominant symmetry. We define the total weight $w = \sum_{\langle i,j\rangle} |\Delta_{ij}|^2$ and calculate the relative weight of each channel as
\begin{equation}
w_\chi = \frac{1}{w} \sum_{\langle i,j\rangle} |\Delta^{\chi}_{ij}|^2,
\end{equation}
By construction, these weights are normalized such that $\sum_\chi w_\chi = 1$.

\subsection*{Linear gap equation and rational pole expansion}
To avoid solving the computationally expensive NLGE we can also identify the leading superconducting instability and its $T_c$ using the linear gap equation (LGE). The LGE is obtained by linearizing the full NLGE, solving for $\Delta_i = U\langle c_{i\uparrow}c_{i\downarrow}\rangle$, around the normal state where $T \approx T_c$ and $\Delta \to 0$. For on-site pairing, this yields
\begin{equation}
    \delta\Delta_{i} = U\sum_{j}\frac{\partial\langle c_{i\uparrow}c_{i\downarrow}\rangle}{\partial\Delta_j}\delta\Delta_j = \sum_j S_{ij}\delta\Delta_j,
\end{equation}
where $S$ is the temperature-dependent pairing stability matrix. A superconducting instability occurs when the leading eigenvalue of $S$, denoted as $\lambda_{\max}$, reaches unity. The pairing temperature $T_c$ is therefore implicitly defined by the condition $\lambda_{\max}(T_c)=1/U$.

The construction of $S$ requires evaluating the linear response of the anomalous correlator, $\langle c_{i\uparrow}c_{i\downarrow}\rangle$, to the superconducting gap $\Delta_i$. To perform this efficiently, we employ a rational pole expansion (RPE) of the Fermi function \cite{moussa2016minimax, lothman2022nematic}. Because the Fermi-Dirac distribution possesses poles in the complex plane, rational approximations are significantly more efficient than standard kernel polynomial expansions for calculating low-temperature superconducting properties \cite{braun2014numerical, moussa2016minimax}. 
For generality, we below derive the response of a non-local correlator $\langle c_{i\uparrow}c_{j\downarrow}\rangle$ with respect to a general pairing term $\Delta_{mn}$ to accommodate for all possible pairing scenarios.
In the presence of a periodic supercell, the stability matrix is also most efficiently evaluated by transforming the correlators to momentum space.  

We start by approximating, within the RPE, the Fermi function at inverse temperature $\beta$ for a Hamiltonian $H(\boldsymbol{k})$ as
\begin{equation}
\label{eq:RPE}
f(H(\boldsymbol{k})) \approx \sum_{p=1}^{N_p} R_p\left(\beta H(\boldsymbol{k}) - P_p\right)^{-1},
\end{equation}
where $P_p$ and $R_p$ denote the $p$-th pole and residue, respectively.
We next introduce useful notation for the BdG Hilbert space. Let $|i\rangle \in \mathcal{H}$ denote a canonical basis vector of the single-particle Hilbert space $\mathcal{H}$, defined such that for a state $|i\rangle$, the $i$-th degree of freedom is one and all other entries are zero. The BdG Hamiltonian acts on $\mathcal{H}_{\text{BdG}} = \mathcal{H} \otimes \mathcal{H}^*$, with particle and hole basis states defined as
\begin{equation}
|i_e\rangle = \begin{pmatrix} |i\rangle \ 0 \end{pmatrix}^T, \quad |i_h\rangle = \begin{pmatrix} 0 \ |i\rangle \end{pmatrix}^T.
\end{equation}
Within this representation, the anomalous correlator for a given momentum is expressed as
\begin{equation}
\begin{aligned}
\langle c_{i\boldsymbol{k}} c_{j-\boldsymbol{k}} \rangle
&=\langle j_e | f(H_{\text{BdG}}(\boldsymbol{k})) | i_h \rangle \\
&\approx \sum_{p=1}^{N_p} R_p \langle j_e | \left(\beta H_{\text{BdG}}(\boldsymbol{k}) - P_p \right)^{-1} | i_h \rangle.
\end{aligned}
\end{equation}

To identify the superconducting transition, we require the static response of the anomalous correlator in the limit $\Delta \to 0$. Using the identity $\partial_\lambda O_\lambda^{-1} = -O_\lambda^{-1} (\partial_\lambda O_\lambda) O_\lambda^{-1}$, for an operator $O$ and parameter $\lambda$, the linear response is given by
\begin{widetext}
\begin{equation}
\label{eq:staticRPE}
\frac{\partial\langle c_{i\boldsymbol{k}}c_{j-\boldsymbol{k}}\rangle}{\partial \Delta_{mn}} \approx -\sum_{p=1}^{N_p}\beta R_p\langle j_e|\left(\beta H_{\text{BdG}}(\boldsymbol{k})-P_p\right)^{-1}\frac{\partial H_{\text{BdG}}(\boldsymbol{k})}{\partial \Delta_{mn}}\left(\beta H_{\text{BdG}}(\boldsymbol{k})-P_p\right)^{-1}|i_h\rangle.
\end{equation}
\end{widetext}
At the critical temperature $T_c$, where $\Delta \to 0$, the BdG Hamiltonian $H_{\text{BdG}}(\boldsymbol{k})$ becomes block-diagonal. In this limit, the derivative $\partial H_{\text{BdG}}/\partial \Delta_{mn}$ simplifies to an operator that encodes the specific pairing symmetry of the system. Under these conditions, Eq.~\eqref{eq:staticRPE} reduces to
\begin{equation}
\frac{\partial \langle c_{i\boldsymbol{k}} c_{j-\boldsymbol{k}} \rangle}{\partial \Delta_{mn}} \approx \sum_{n=1}^{N_p} \beta R_p \langle y_{j\boldsymbol{k}}^p | \frac{\partial \Delta(\boldsymbol{k})}{\partial \Delta_{mn}} | x_{i\boldsymbol{k}}^p \rangle,
\label{eq:rpelge}
\end{equation}
where the vectors $|x_{i\boldsymbol{k}}^p\rangle$ and $|y_{j\boldsymbol{k}}^p\rangle$ are obtained by solving
\begin{equation}
(\beta H_0(\boldsymbol{k}) - P_p) |x_{i\boldsymbol{k}}^p\rangle = |i\rangle, \qquad (\beta {H}_0^*(-\boldsymbol{k}) + {P}^*_p) |y_{j\boldsymbol{k}}^p\rangle = |j\rangle.
\end{equation} 
Equation~\eqref{eq:rpelge} expresses the kernel of the linearized gap equation in terms of solutions to linear systems that can be computed independently for each pole and momentum, allowing for highly efficient parallelization. 

In this work, we determine the critical on-site interaction strength $U$ at which the pairing instability first occurs, for several chosen $T_c$. We again align the Fermi level with the disored-induced DOS peak. For vacancies, the uniform density theorem \cite{lieb1993uniform} again ensures we need not have to deal with the Hartree shift. In the case of hydrogenation, we do not solve for $\mu$ self-consistently due to an increased computational complexity, but rather fix it by hand such that the disorder-induced DOS peak still remains aligned with the Fermi energy. Consequently, hydrogenation requires some tunable doping, through e.g.~electrostatic gating, to reach the reported $T_c$.

\subsection*{Superconducting phase coherence}
\subsubsection*{Superfluid weight}
To establish superconducting phase coherence, we compute the  superfluid weight $D^s_{\mu\nu}$, along spatial coordinates $\mu,\nu$, using linear-response theory. For the derivation below we focus on the case of on-site pairing. For a more thorough derivation, including expressions for nearest-neighbor pairing, we refer to \cite{huhtinen2022revisiting, lamponen2025superconductivity}. The total superfluid weight is given by \cite{huhtinen2022revisiting, lamponen2025superconductivity}
\begin{widetext}
\begin{equation}
\begin{aligned}
\label{eq:sfwfull}
D^{s}_{\mu \nu} = \frac{1}{V} \sum_{\boldsymbol{k}}\sum_{mn}
\frac{f(E_{m\boldsymbol{k}})-f(E_{n\boldsymbol{k}})}{E_{n\boldsymbol{k}}-E_{m\boldsymbol{k}}}
\Big[
&\langle \psi^n_{\boldsymbol{k}} | \partial_\mu H_{\text{BdG}} | \psi^m_{\boldsymbol{k}} \rangle
\langle \psi^m_{\boldsymbol{k}} | \partial_\nu H_{\text{BdG}} | \psi^n_{\boldsymbol{k}} \rangle- \\
&\langle \psi^n_{\boldsymbol{k}} | \partial_\mu H_{\text{BdG}}\gamma^z | \psi^m_{\boldsymbol{k}} \rangle
\langle \psi^m_{\boldsymbol{k}} | \partial_\nu H_{\text{BdG}}\gamma^z | \psi^n_{\boldsymbol{k}} \rangle
\Big].
\end{aligned}
\end{equation}
\end{widetext}
Here, $\gamma^z = \sigma_z \otimes I$ is the Pauli operator in particle-hole space and $V$ is the volume of the unit cell. In the limit $E_{n\boldsymbol{k}} \to E_{m\boldsymbol{k}}$, the prefactor reduces to $-\partial_E n_F(E)|_{E=E_{n\boldsymbol{k}}}$.

To highlight the role of quantum geometry, we decompose $D^s_{\mu\nu}$ into its conventional and geometric contributions by expanding the BdG eigenstates in terms of the normal-state Bloch basis $|m_{\boldsymbol{k}}\rangle$. The BdG eigenstates are then decomposed as
\begin{equation}
\label{eq:proj}
    |\psi^n_{\boldsymbol{k}}\rangle = \sum_m \left( \alpha_{nm} |+\rangle \otimes |m_{\boldsymbol{k}}\rangle + \beta_{nm} |-\rangle \otimes |m^*_{-\boldsymbol{k}}\rangle \right),
\end{equation}
where $|\pm\rangle$ denote the eigenvectors of $\sigma_z$ in particle-hole space with eigenvalues $\pm 1$, and $\alpha_{nm}$ and $\beta_{nm}$ represent the quasiparticle amplitudes in the normal-state Bloch basis. By substituting this expansion into Eq.~(\ref{eq:sfwfull}) and introducing the normal-state current operators
\begin{equation}
\begin{aligned}
    \label{eq:currentop}
    \left[j_{\mu}(\boldsymbol{k})\right]_{m n}&=\langle m_{\boldsymbol{k}}| \partial_{\mu} H(\boldsymbol{k})|n_{\boldsymbol{k}}\rangle\\
    &=\partial_{\mu} \varepsilon_{m} \delta_{m n}+\left(\varepsilon_{m}-\varepsilon_{n}\right)\left\langle\partial_{\mu} m_{\boldsymbol{k}} |n_{\boldsymbol{k}}\right\rangle
    \end{aligned}
\end{equation}
along with the prefactors
\begin{equation}
    \label{eq:prefactors}
    C^{mn}_{pq} = -2\sum_{ij} \frac{f(E_j) - f(E_i)}{E_i-E_j} \alpha_{im}^* \alpha_{jn} \beta_{jp}^* \beta_{iq},
\end{equation}
we arrive at an alternative expression for the superfluid weight
\begin{equation}
\begin{aligned}
    \label{eq:fullsfw3}
    D^s_{\mu\nu} =& \frac{1}{V}\sum_{\boldsymbol{k}} \sum_{mn} \sum_{pq}C^{mn}_{pq}\Big(\left[j_{\mu}(\boldsymbol{k})\right]_{mn}\left[j_{\nu}(\boldsymbol{-k})\right]_{qp} 
    \\+& \left[j_{\nu}(\boldsymbol{k})\right]_{mn}\left[j_{\mu}(\boldsymbol{-k})\right]_{qp} \Big).
    \end{aligned}
\end{equation}
This expression naturally separates into an intraband (conventional) contribution $D^s_{\mu\nu,\text{conv}}$, determined by the diagonal elements of the current operator, and an interband (geometric) contribution $D^s_{\mu\nu,\text{geom}}$, which depends on the off-diagonal elements and encodes the quantum geometry of the bands. For isolated flat bands, the quantum geometrical contribution is directly proportional to the normal-state quantum metric \cite{peotta2015superfluidity, liang2017band, huhtinen2022revisiting}.

Direct evaluation of Eq.~\eqref{eq:fullsfw3} is computationally prohibitive due to the multiple nested sums. Consequently, we compute the total superfluid weight using Eq.~\eqref{eq:sfwfull}, while the conventional contribution is evaluated separately via the diagonal elements of the current operator. The geometric contribution is then determined as the difference $D^s_{\mu\nu,\text{geom}} = D^s_{\mu\nu} - D^s_{\mu\nu,\text{conv}}$.

\subsubsection*{BKT Temperature}
To determine the BKT transition temperature $T_{\text{BKT}}$, which marks the onset of phase coherence, we solve the NLGE self-consistently at temperatures below $T_c$, evaluate the superfluid weight, and iterate until the Nelson–Kosterlitz criterion
\begin{equation}
\label{eq:nk}
T_{\text{BKT}} = \frac{\pi}{8}\sqrt{\det D^s(T_{\text{BKT}})}
\end{equation}
is satisfied within an absolute tolerance of $5\cdot10^{-4}\,\text{eV}$. We employ Brent’s method to solve the resulting nonlinear root-finding problem efficiently \cite{brent1973some}.

\subsubsection*{Quantum metric}
To understand the quantum geometric contribution to the superfluid weight, we compute the quantum metric of the normal state, which is the real part of the quantum geometric tensor (QGT), $g_{\mu\nu} = \mathrm{Re}\, \mathcal{Q}_{\mu\nu}$ \cite{peotta2015superfluidity}. For a set of occupied bands $\mathcal{B}$, the (integrated) QGT is defined as
\begin{equation}
    \label{eq:QGT}
    \mathcal{Q}_{\mu\nu} = \int_{\text{BZ}} \text{d}\boldsymbol{k} \sum_{n\in\mathcal{B}}\sum_{m\notin\mathcal{B}}
    \langle \partial_{\mu}\varphi^n_{\boldsymbol{k}} \mid \varphi^m_{\boldsymbol{k}} \rangle
    \langle \varphi^m_{\boldsymbol{k}} \mid \partial_{\nu}\varphi^n_{\boldsymbol{k}} \rangle,
\end{equation}
where the integral is over the first supercell BZ and $|\varphi^n_{\boldsymbol{k}}\rangle$ denote the eigenstates of the normal-state Hamiltonian $H_0$.
Using standard perturbation theory, the derivatives of the eigenstates can be expressed in terms of derivatives of $H_0$ and its eigenenergies $\xi_{n\boldsymbol{k}}$, yielding
\begin{equation}
    \label{eq:QGT2}
    \mathcal{Q}_{\mu\nu} = \int_{\text{BZ}}\text{d}\boldsymbol{k}\sum_{n\in\mathcal{B}}\sum_{m\notin\mathcal{B}} \frac{\langle\varphi^n_{\boldsymbol{k}}|\partial_{\mu}H_0|\varphi^m_{\boldsymbol{k}}\rangle\langle\varphi^m_{\boldsymbol{k}}|\partial_{\nu}H_0|\varphi^n_{\boldsymbol{k}}\rangle}{(\xi_{n\boldsymbol{k}}-\xi_{m\boldsymbol{k}})^2}.
\end{equation}
We focus on the trace of $g$, as it is the simplest basis-invariant scalar characterizing the quantum geometry.

Due to the embedding dependence of the quantum metric, the orbital positions in the unit cell influence the magnitude of the total superfluid weight through the geometrical contribution \cite{huhtinen2022revisiting}. A consistent evaluation of the superfluid weight, therefore, requires fixing a representation in which the quantum metric is minimized \cite{huhtinen2022revisiting}. To minimize the quantum metric, we adopt a variational approach and perform a line search over orbital positions, selecting those that minimize the trace of the quantum metric (Extended Data Fig.~\ref{fig:s4}). We find that the standard atomic orbital positions of the honeycomb lattice minimize the quantum metric across all vacancy concentrations considered.

\subsection*{Singularity spectrum}
Multifractality can be understood through the singularity spectrum $f(\alpha)$, which we calculate using the method of Chhabra and Jensen \cite{chhabra1989direct}. Below, we restrict the derivation for the magnitude of the on-site order parameter, but a similar derivation is also possible for unconventional pairing.

We start by defining a probability measure on the disordered graphene lattice of size $L$
\begin{equation}
    \lambda_i=\frac{|\Delta_i|^2}{\sum_j|\Delta_j|^2}.
\end{equation}
The lattice is then partitioned into domains of size $\ell$, and the coarse-grained probability measure in each domain is computed as
\begin{equation}
    \lambda_k(\ell) = \sum_{i\in k}\lambda_i.
\end{equation}
For each moment $q$, we define normalized weights
\begin{equation}
    w_k(q,\ell) = \frac{\lambda_k(\ell)^q}{\sum_{k^\prime}\lambda_{k{^\prime}}(\ell)^q}.
\end{equation}
We further compute the quantities
\begin{equation}
    \begin{aligned}
    A_q(\ell) &= \sum_k w_k(q,\ell)\ln \lambda_k(\ell), \\
    F_q(\ell) &= \sum_k w_k(q,\ell)\ln w_k(q,\ell),
    \end{aligned}
\end{equation}
from which the singularity strength $\alpha$ and spectrum $f(\alpha)$ are extracted as the slopes of the linear fits of $A_q(\ell)$ and $F_q(\ell)$, respectively, as functions of $\ln(\ell/L)$.

\section*{Data availability}
The data that support the findings of this work are openly available \cite{zenodo}.
\section*{Code availability}
Code is available from the authors upon reasonable request.
\section*{Acknowledgments}
We thank L.~Baldo, K.-E.~Huhtinen, and Q.~Marsal for valuable discussions. This work was supported by the Swedish Research Council (Vetenskapsr\aa det) Grant No.~2022-03963 and the European Research Council (ERC) under the European Union’s Horizon 2020 research and innovation programme (ERC-2022-CoG, Grant agreement No.~101087096). Views and opinions expressed are, however, those of the author(s) only and do not necessarily reflect those of the European Union or the European Research Council Executive Agency. Neither the European Union nor the granting authority can be held responsible for them. The calculations were enabled by resources provided by the National Academic Infrastructure for Supercomputing in Sweden (NAISS), partially funded by the Swedish Research Council through grant agreement No.~2022-06725. We further acknowledge NAISS for awarding this project access to the LUMI supercomputer, owned by the EuroHPC Joint Undertaking and hosted by CSC (Finland) and the LUMI consortium.
\section*{Author Contributions}
TL and ABS conceived the idea. TL performed preliminary calculations and reported some of the main findings. JvP performed all calculations reported in the manuscript, and JvP and ABS analyzed the results. JvP wrote the first draft of the manuscript, with JvP and ABS revising the manuscript. ABS supervised the project. 

\section*{Competing interests}
The authors declare no competing interests.
\appendix

\pagebreak
\clearpage

\onecolumngrid
\section*{Extended data for ``Disorder-induced superconductivity in graphene"}
\setcounter{figure}{0}
\begin{figure*}[ht!]
    \centering
    \includegraphics[width=0.9\linewidth]{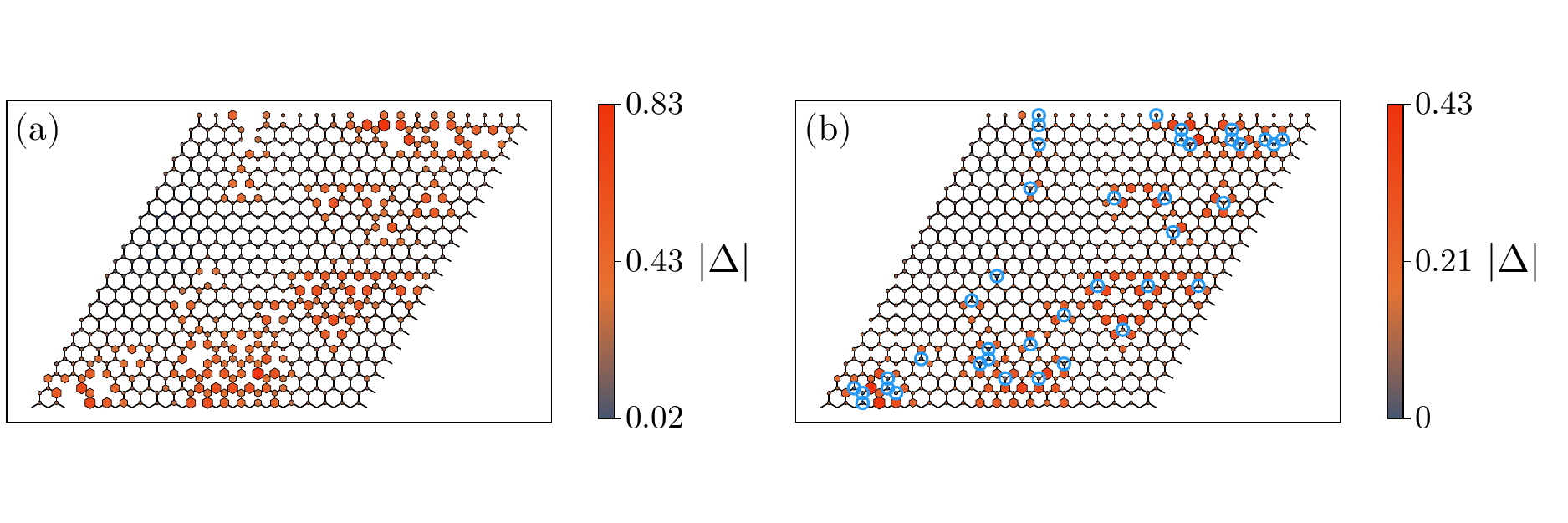}
    \caption{\textbf{Conventional superconducting order parameter.} Spatially resolved on-site superconducting order parameter at on-site interaction strength $U=0.67\,U_c$ and disorder concentration $n=4\,\%$ for vacancies (a) and hydrogenation (b). Hydrogenated sites in (b) are circled in blue.}
    \label{fig:s1}
\end{figure*}
\begin{figure*}[ht!]
    \centering
    \includegraphics[width=0.9\linewidth]{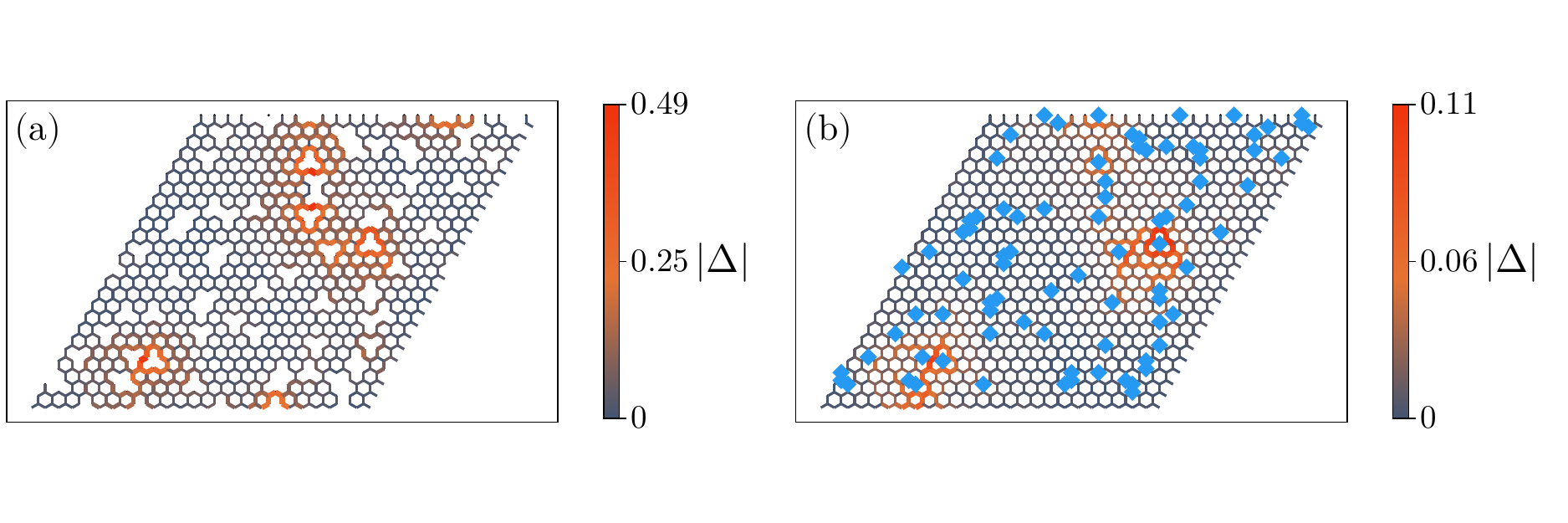}
    \caption{\textbf{Unconventional superconducting order parameter.} Spatially resolved unconventional superconducting order parameter at nearest-neighbor interaction strength $J= 0.63\,J_c$ and disorder concentration $n=6\,\%$ for vacancies (a) and hydrogenation (b). Hydrogenated sites in (b) are circled in blue.}
    \label{fig:s2}
\end{figure*}

\begin{figure*}[ht!]
    \centering
    \includegraphics[width=0.9\linewidth]{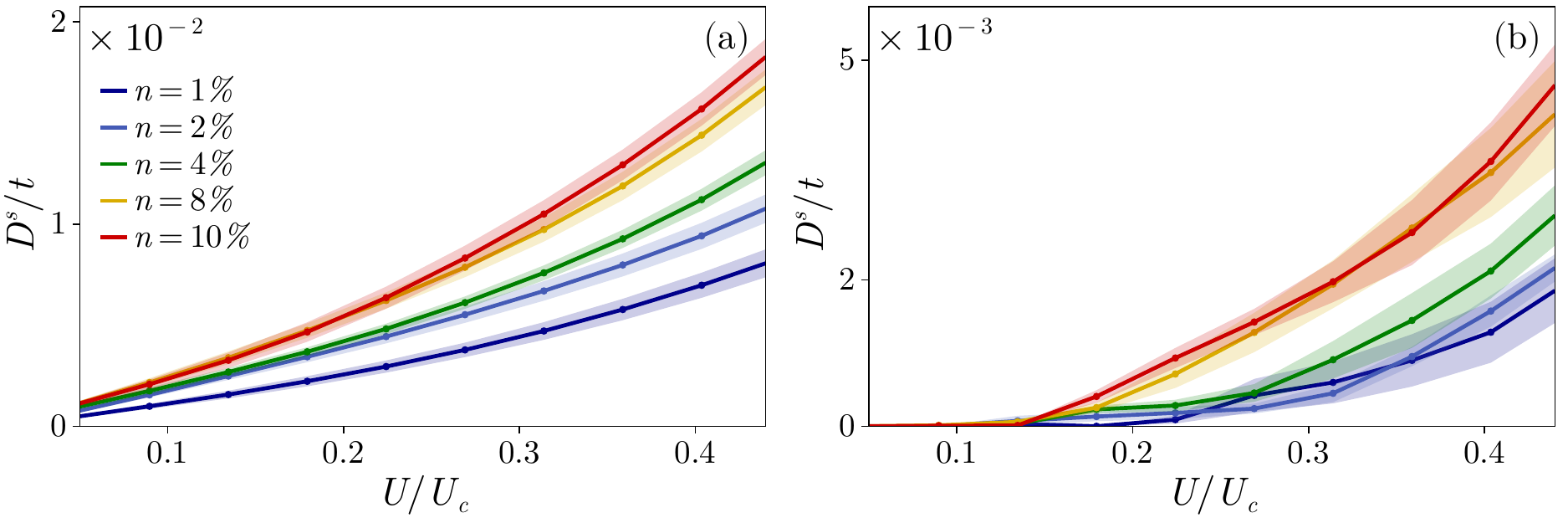}
    \caption{\textbf{Zero-temperature superfluid weight.} Superfluid weight as a function of on-site attractive interaction $U$, normalized to the critical $U_c$ for superconducting pairing in pristine graphene, for a range of disorder concentrations $n$ of vacancies (a) and hydrogenation (b). Results are averaged over five disorder configurations. Shaded regions indicate standard deviation of the mean.}
    \label{fig:s3}
\end{figure*}

\begin{figure*}[ht!]
    \centering
    \includegraphics[width=0.6\linewidth]{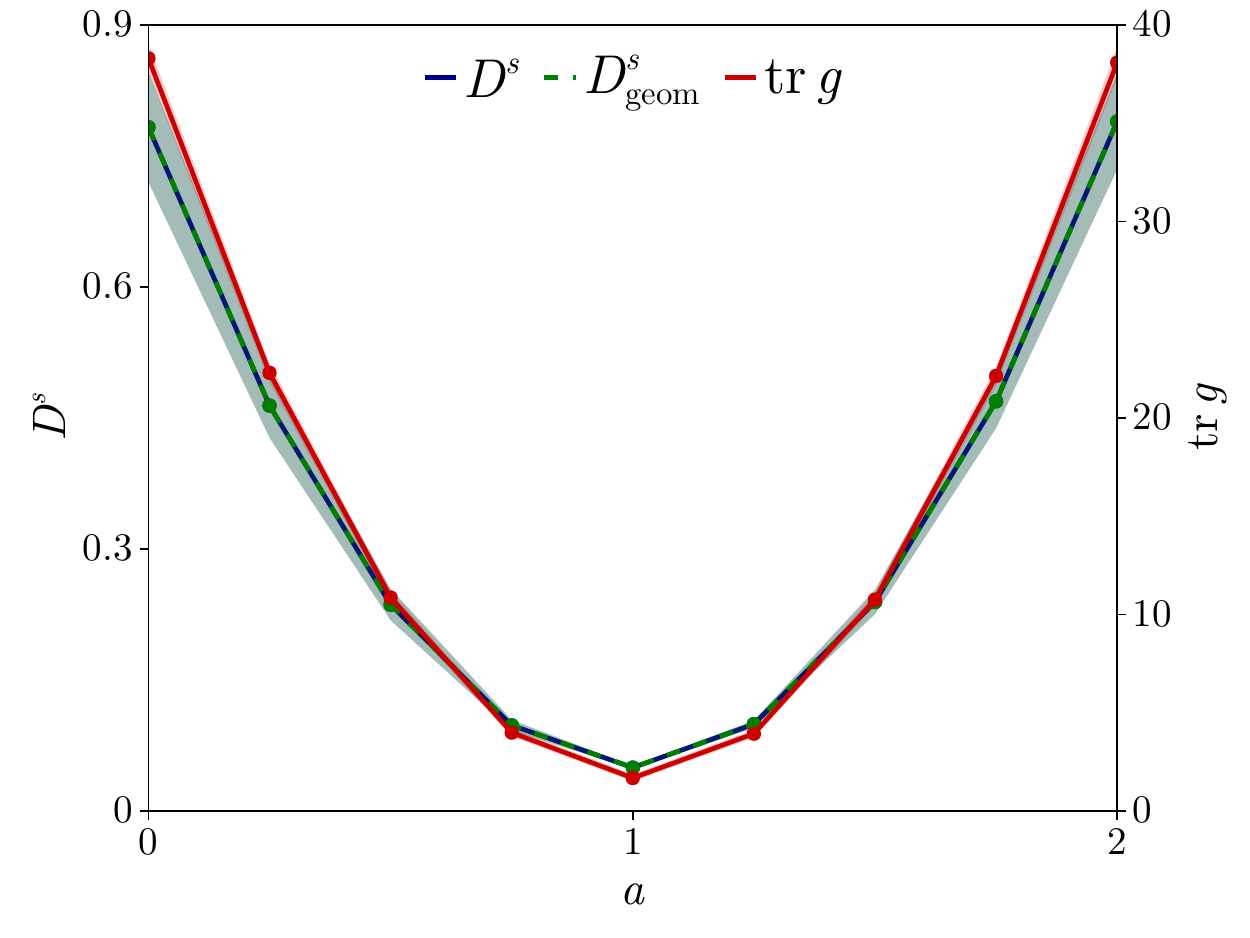}
    \caption{\textbf{Minimal quantum metric.} Total superfluid weight $D^s$, geometric $D^s_{\rm geom}$ contribution, and trace of the quantum metric tr $g$ as a function of orbital positions $a$ for interaction strength $U = 0.45 \,U_c$ and vacancy concentration $n=10\,\%$.  For $a=0$, all orbitals overlap, $a=1$ recovers the original graphene unit cell, and intermediate values of $a$ scale the geometry continuously. Results are averaged over five disorder configurations. Smaller disorder concentrations behave similarly.}
    \label{fig:s4}
\end{figure*}


\end{document}